\RequirePackage{fix-cm}
\documentclass[twocolumn,epjc3]{svjour3}
\smartqed  % flush right qed marks, e.g. at end of proof
\RequirePackage{graphicx}

\newcommand{\p}{\partial}
\newcommand{\sh}{\sinh}
\newcommand{\eqref}[1]{(\ref{#1})}
\newcommand{\eq}[2]{\begin{equation}\label{#2}
	#1
\end{equation}}
\newcommand{\cf}[2]{\Gamma^{#1}_{#2}}
\newcommand{\etal}{et al.}

\journalname{Eur. Phys. J. C}
\begin{document}

\title{Development of Zeldovich's approach for cosmological distances measurement in the Friedmann Universe}

\titlerunning{The cosmological distances measurement in the Friedmann Universe}        % if too long for running head

\author{A.V. Nikolaev\thanksref{e1,addr1}
        \and
        S.V. Chervon\thanksref{e2,addr1,addr2} %etc.
}

\thankstext{e1}{e-mail: ilc@xhns.org}
\thankstext{e2}{e-mail: chervon.sergey@gmail.com}


\institute{Ilya Ulyanov State Pedagogical University, 100 years of V.I. Lenin's Birthday Square, B. 4, 432700 Ulyanovsk, Russia \label{addr1}
           \and
 Astrophysics and Cosmology Research Unit,
School of Mathematics, Statistics and Computer Science,
University of KwaZulu-Natal, Private Bag X54 001,
Durban 4000, South Africa
           \label{addr2}
           %\and
           %\emph{Present Address:} if needed\label{addr3}
}

\date{Received: date / Accepted: date}

\maketitle

\begin{abstract}

We present our development of Ya. Zeldovich's ideas for the measurement of the cosmological angular diameter distance (ADD) in the Friedmann Universe.
We derive the general differential equation for the ADD measurement which is valid for an open, spatially-flat and closed universe, and for any stress energy tensor.
We solve these equations in terms of quadratures in a form suitable for further numerical investigations for the present universe filled by radiation, (baryonic and dark) matter and dark energy.
We perform the numerical investigation in the absence of radiation, and show the strong dependence ADD has on the filling of the cone of light rays (CLR).
The difference of the empty and totally filled CLR may reach 600-700 Mps. for a redshift of $f\simeq 3$.
\end{abstract}

\section{Introduction}
\label{intro}
In the present article, we are going to reconsider the issue of cosmological distances measurement in cosmology.
Methods commonly used by astronomers are collected in the well cited review \cite{Hogg_2000} and it may be considered a good introduction to this topic.

The angular diameter distance and the luminosity distance are known for being of most practical use.

These two distances are connected by the following relation \cite{Ellis_2007}:
\eq{
d_l = (1+f)^2d_a
}{d_l2d_a}

where $d_l$ is the luminosity distance, $d_a$ the angular diameter distance, and $f$ the redshift. This relation gives us a feasibility to concentrate our study on the angular diameter distance.

As a rule, the derivation of the angular diameter distance \cite{Hogg_2000} is undertaken for the homogeneous Universe \cite{Weinberg_2008}, i.e. all matter of the Universe is distributed homogeneously by the assumption.
This assumption is valid for volumes of the order of about 500 Mps as the side of a cube, and it does not reflect the real situation in the case of the distance measurement when the beam of light is propagating through a generally small volume.

\begin{figure}[!htb]
\centering
\includegraphics[width=8cm]{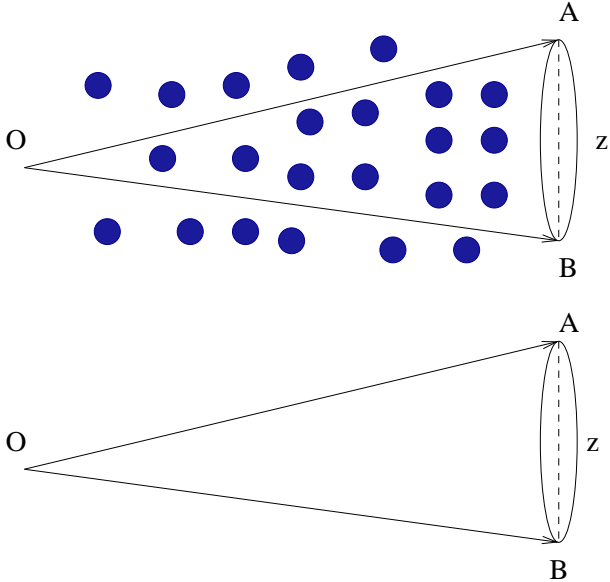}
\caption{The angular diameter distance setup. Field and empty cone of light rays schemes.}
\label{figure1}
\end{figure}

Let us briefly define the problem of cosmological angular diameter distance (ADD) measurement. The definition of ADD, which is valid in Euclidean space, is usually extended to a curved space \cite{Zeldovich_1971} with the formula
\eq{
d_a = \frac{z}{\phi}
}{d_a_def}
Here, $z$ is the linear size of the object and $\phi$ is its angular size (Figure \ref{figure1}). In the Friedmann-Robertson-Walker geometry %(\ref{metric})
we can find the linear size of a distant object using the metric
\eq{ds^2=
dt^2-a(t)^2\left( \frac{dr^2}{1-kr^2}+ r^2(d\theta^2 + \sin^2\theta d\phi^2)\right).
}{frw-1}

For example, in the spatially-flat universe, where $ k=0$, one can obtain
\eq{
z = a_er\phi
}{zflat}
where $a_e$ is the scale factor at the time of emission. Combining the expressions above, it is easy to obtain (commonly used by astronomers) the  equation for the angular diameter distance:
\eq{
d_a = a_er
}{d_aflat}

To obtain the result (\ref{d_aflat}) we used the FLRW metric (\ref{frw-1}) corresponding to the homogeneous universe. Therefore, if we suggested that our universe were not homogeneous in the cone of light rays (CLR)\footnote{In the range with the term "the cone of light rays"\ we will also use the term "a light cone" as the cone of null geodesics congruence.} of ADD measurement, then we must take into account local inhomogeneities inside or close to the CLR.

What will the ADD measurement  be if we take into account a local inhomogeneity?
As far as we know, the first man who dealt with this question was Ya. Zeldovich. He presented the solution for "the homogeneous in the mean Universe" in \cite{Zeldovich_1964}.
It is interesting to mention that in the monograph
\cite{Zeldovich_1971} I. Novikov and Ya. Zeldovich
noted that during the Symposium in Burokan (1966), American astronomers reported that familiar ideas were declared by R. Feinman.
Recently in the paper \cite{Clarkson_2012} it was also confirmed that R. Feinman pondered over the same problem, and also suggested that one should consider a zero stress-energy tensor inside a light cone.

Considering a congruence of null geodesics, we will use the following terminology proposed by R. Penrose  \cite{Penrose_1966}:
\begin{itemize}
	\item The Ricci focusing is the focusing due to the gravitational effect of intrinsic mass inside the light cone;
	\item The Weil focusing is the focusing due to inhomogeneous clumps of matter along the null geodesics path.
\end{itemize}

Let us review the main stream of Zeldovich's original ideas represented in the papers \cite{Zeldovich_1964},\cite{Zeldovich_1965}.
In the article \cite{Zeldovich_1964} Ya. Zeldovich introduced "a homogeneous in the mean universe" and he analyzed the effect of the local non-uniformity of the matter-dominated spatially-flat Friedmann Universe on the angular and luminosity distances measurement.
It was found for the ADD under suggestion that there was a negligible amount of matter inside the light cone and it was possible to neglect the gravitational effect of that matter.
The method applied is the integration of numerous lensing deflections due to the intrinsic mass of a light cone.

The solutions of ADD measurement for a nonhomogeneous nonflat universe were found in the paper by V. Dashevskii and Ya. Zeldovich \cite{Zeldovich_1965}. In this paper they also presented the method for describing Ricci focusing through momentums of a photon. Later on we will explain this approach in detail where we make use of this method in our study.
In the work by V. Dashevskii and V. Slysh \cite{Dashevskii_1965} the analytical solution for a closed matter-dominated Universe for a partly filled light cone was found. They also derived the differential equation on the linear distance between the two rays emitted by  the outer points of the object. Let us mention here that this equation does not contain the dark energy component and it may not be applied for ADD measurement now from the current observational data.

Let us note that Zeldovich's original ideas were used afterwards in a series of papers by C. Dyer and R. Roeder \cite{Dyer_1972}, \cite{Dyer_1974}, \cite{Dyer_1973}.
Their papers are well cited and we comment on them briefly.

In the first paper \cite{Dyer_1972} they used Sachs' equations \cite{Sachs_1961} and obtained the result which was found in \cite{Zeldovich_1964}. In the second paper \cite{Dyer_1973} they obtained a differential equation similar to the equation from \cite{Dashevskii_1965} which was valid only for the Universe without dark energy, because the energy momentum tensor was
set up as $T_{\mu\nu}=diag(\rho, 0,0,0)$. The solution of this equation proves the result from \cite{Dashevskii_1965}. In the third paper \cite{Dyer_1974} the result was found for the Swiss Cheese Universe.

In 1976, S. Weinberg \cite{Weinberg_1976} showed that the summation of gravitational deflection caused by individual clumps of matter is equal to the effect caused by the homogeneous distribution of the same mass. This paper narrowed the interest of the community to effects of inhomogeneity to the distance measurement and
it provides a strong criticism of the Swiss Cheese Universe model. Nonetheless, S. Weinberg's results are in agreement with Ya. Zeldovich's ideas. Indeed, the matter enclosed inside a light cone for the homogeneous in the mean Universe may be considered as a homogeneous distribution (of a small density).
What is the problem with the direct application of the original solutions of Ya. Zeldovich for calculating cosmological distances? The problem is due to the fact that these solutions were obtained in the absence of dark energy and dark matter, i.e. for the Universe which contains only baryonic matter. This is controversial in our current understanding of the Universe. Therefore in the present paper we are going to extend Ya. Zeldovich's original ideas for solving this problem in a general form.

Let us mention the series of papers by C. Alcock and N. Anderson where they discuss the problem of distance measurement. In the first paper \cite{Alcock_1984} they emphasized the importance of  correct cosmological distance measurement for calculating the Hubble constant through gravitational lensing. In the second paper \cite{Alcock_1985} they presented an original method for the distance measurement; the so called "effective distance".
The problem with their approach is that all our cosmological and astronomical theories were developed for angular diameter (luminosity) distance. This is why we are not ready to apply "effective distance"\ methods and we are going to keep commonly accepted definitions of fundamental concepts.

The ideas of C. Dyer and R. Roeder were developed by R. Kantowski \cite{Kantowski_1998}, who constructed an analysis of differential equations and their solutions for the Swiss Cheese Universe. In the paper \cite{Kantowski_2000} the solutions for various inhomogeneous cosmological models were found. %
We would like to mention the impact of P. Schneider and A. Weiss \cite{Schneider_1988}, S. Seitz \etal \cite{Schneider_1994} in the development of C. Dyer and R. Roeder's ideas.
%papers  \cite{Schneider_1988}, \cite{Schneider_1994}. %

Why should we reconsider previous results in the distance measurements? From our point of view, the main problem in these results is that they were obtained for an inhomogeneous Universe.
Our understanding is that this assumption is too strong, because it is already proven that global Universe mass distribution is homogeneous. Thus we assume that the Universe evolves like a homogeneous universe, but if we measure the distance to an object in the space, the effect of Ricci focusing becomes weaker. The reason is in fact that the density inside the light cone is smaller than the critical density of the Universe. Summing up the discussion above we can state %
that the Ya. Zeldovich model of the Universe (the homogeneous in the mean Universe) is very suitable for describing observations in the present Universe.

Let us explain our last thesis in detail by firstly discussing the method of C. Dyer and R. Roeder. They start from Sachs' equations \cite{Sachs_1961}, which are a version of the Raychaudhuri equation for null geodesics \cite{Poisson_2004}. It should be noted that the Raychaudhuri equation is more general than the Friedmann equation \cite{Ellis_1997} and it already includes the effects of gravitational lensing.
Starting from these mentioned works by C. Dyer and R. Roeder, physicists are using (proposed by R. Penrose \cite{Penrose_1966}) the Weil and Ricci focusing for calculating ADD.

For the Friedmann Universe, the Weil tensor is equal to zero ($W_{iklm} = 0$)  \cite{Ellis_1997}. In the case of small clumps of the Swiss Cheese Universe, C. Dyer and R. Roeder showed \cite{Dyer_1981} that effects of the Weil focusing on every clump can be approximated by the Ricci focusing. This proposition proves S. Weinberg's results \cite{Weinberg_1976}. If we follow Ya. Zeldovich's ideas for the ADD measurement, then our universe remains a Friedmann Universe
and there are no concentrated clumps of matter on the line of sight. Thus the effect of the Weil focusing can be neglected. To take into account the concentrated clumps of matter, we suggest that one should use the well developed gravitational lensing theory. How will we hold the value of the Ricci focusing in cases of the ADD %
measurement? We will follow the Ya. Zeldovich and V. Dashevskii ideas which were published in \cite{Zeldovich_1965}, i.e. we use the fraction of perpendicular and longitudinal components of the photon.

It should be noted that the interest of the scientific community with regards to the problem of distance measurement in an inhomogeneous Universe does not end. For example, in the paper by K. Bolejko  and P. Ferreira \cite{Bolejko_2012}, the authors once again emphasise the importance of the effects of inhomogeneities in cosmology. We should also mention that our idea of developing  Ya. Zeldovich's original ideas for cosmological distance measurement is not new.
In  the paper by R. Kayser \etal   \cite{Kayser_1996} the differential equation presented in the paper \cite{Dashevskii_1965} was used with the aim of finding the cosmological distance in the Universe filled with dark energy. Unfortunately, as it was mentioned before, the differential equation obtained by V. Dashevskii and V. Slysh for the matter-dominated Universe did not account for dark energy and could not be applied to the $\Lambda CDM$ model.
It is easy to check by using standard cosmological parameters and assuming a filled light cone, that the equations obtained by R. Kayser \etal \cite{Kayser_1996} give wrong results in this case.

%insertion 1
The investigation of inhomogeneities and its connection to a luminosity distance (LD) has been performed within the framework of Lemaitre-Tolman-Bondy (LTB) solutions in the work by A. Romano et al \cite{Romano_2012}. It was shown there that an inhomogeneous isotropic universe described by an LTB solution admits a positive, averaged acceleration. Thus, this model may be considered as an alternative to standard FLRW cosmology with dark energy (DE). Also, the effect of inhomogeneities in the presence of a cosmological constant has been considered for LTB solutions which were only locally inhomogeneous. Finally, in respect of LD, it was found that the luminosity distance as a function of redshift  $D_L(z)$ is not significantly affected by small inhomogeneities, but the apparent cosmological observables, derived from $D_L(z)$ under the assumption of homogeneity, are significantly affected because they are sensitive to its derivatives.  Further investigation of the effects of primordial curvature perturbations on the apparent value of a cosmological constant, using the LTB solution, was performed in the work by A. Romano et al \cite{Romano_2014}.

%insertion 2
Series of works (\cite{Gasperini_2011}; \cite{Ben-Dayan_2013}; \cite{Ben-Dayan_2012}; \cite{Ben-Dayan_2013_2}; \cite{Fanizza_2013}; \cite{Ben-Dayan_2014}) are devoted to investigation of the luminosity-redshift relation up to second order in perturbation theory using a very promising geodesic light-cone (GLC) gauge, first proposed in the work by M. Gaspirini et al \cite{Gasperini_2011}. The LCG approach is based on null geodesics as well and is present in our work. The key differences in the methods are a consideration of a perturbed FLRW metric in LCG and an analysis of homogeneous in the mean universe used in our approach. The luminosity distance is known to be computed to first order in the longitudinal gauge in the works by M. Sasaki \cite{Sasaki_1987} and M. Kasai \& M. Sasaki \cite{Kasai_1987}. For example, in the recent work by G. Marozzi \cite{Marozzi_2015} the expressions for the redshift and luminosity distance--redshift relation in a generic homogeneous FLRW universe with anisotropic stress have been computed with perturbation up to second order.

%insertion 3
The authors of the work by O. Umeh et al. \cite{Umeh_2014} noted that precision cosmology from the next generation of telescopes should be complemented with theoretical models of the same level of precision. To this end, the distance--redshift relation in \cite{Umeh_2014} was extended from first order to second order in cosmological perturbation theory for a general dark energy model. In \cite{Umeh_2014_2} the derivation of the distance--redshift relation was presented in detail and the observed redshift and the lensing magnification to second order in perturbation theory was found. Let us note that in \cite{Umeh_2014},\cite{Umeh_2014_2} the ADD and luminosity distance of a source are considered the same, i.e. they are connected with the Etherington identity\cite{Ellis_2007}.

%insertion 4
We should stress here, that in the present work we are considering the angle diameter distance in cosmology. It is well known that for ADD measurements a very small angle (of a few arc seconds) is under consideration. For this reason, we cannot tell about a homogeneous and isotropic universe in this thin cone of light rays and thus the use of the FLRW metric is questionable in this case. When the luminosity distance measurement is under consideration, we use the total celestial sphere, which includes the properties of homogeneity and isotropy so that the application of the FLRW metric is justified. From this position, we can tell that the Etherington identity may have small deviations for large redshifts \cite{Nikolaev_2015_2}.
%endof insetions

The paper is organized as follows: in section 2 we present the general approach for the ADD measurement in the Friedmann Universe and we derive the equation for evolution of the cross-section diameter $z$ of a light beam from a distance object. We show that the derived equation is valid for any type of Friedmann Universe, for any forms of energy momentum tensor and the equation includes the separate term which is responsible for Ricci focusing.
In section 3, we apply the general formula obtained in section 2 for $z$ for ADD calculation in the Friedmann Universe of any type, filled by dark energy and radiation. We demonstrate also that the formula for ADD calculation is valid for a wide range of modified gravity theories.
We present the numerical solution for a partly filled light cone in section 4.

\section{ADD in the Friedmann Universe: general approach}
\label{sec:1}
%\section{ADD in the Friedmann Universe: general approach}

Let us start from the definition of the angular diameter distance \eqref{d_a_def} which is generally exploited in astronomy. As mentioned before, it is the fraction of linear and angular sizes of the object. In astronomy we generally deal with spherical bodies and therefore we can simpify our task by considering the diameter of the object ($z$ in \eqref{d_a_def}) instead of its area. As we are discussing a congruence from a distance object of radial null geodesics which are crossed at a point of observation (center of our spherical coordinate system), the angle $\phi$ between the boundary points of a diameter $z$ remains constant by its definition \eqref{d_a_def}.

Let us choose a spherical coordinate system $[t,r,\theta,\phi]$ with an observer placed in the center of it.
To calculate the changing of the linear size $z$ during the travel of the light beams, we have to account for the following effects:

\begin{enumerate}
\item Expansion of the universe.
\item The Ricci focusing.
\end{enumerate}

Let us remind ourselves that in the Friedmann Universe all matter is involved with the Hubble current and in the comoving coordinates (with the Earth observer in the center), the spatial coordinates of all particles are not changing $\dot{x}_i = 0$ \cite{applebook}. Therefore we have to consider not the object itself, but the photons which are moving along null radial geodesics. In this context $z$ will be the distance between two light rays from the end points of the object at the time of emission. Evidently $z$ will be variable.

Our next task is to derive the differential equation involving $z$ which allows us to take into account the effects arising when measuring distances in a Friedmann Universe.

For the Friedmann-Lemaitre-Robertson-Walker (FLRW) metric, we will use two forms

\eq{
ds^2 = dt^2 - a^2[dr^2 + f^2(r)(d\theta^2 + \sin^2\theta d\phi^2)]
}{123}
\eq{
=dt^2-a(t)^2\left( \frac{dr^2}{1-kr^2}+ r^2(d\theta^2 + \sin^2\theta d\phi^2)\right)
}{metric}
where $f(r)= \sin r, r , \sinh r $ or k=1,0,-1 for a closed, spatially-flat, open universe respectively.
For the sake of simplicity we choose $\theta = \frac{\pi}{2}$.
Let us analyze, with radial null geodesics  $d\phi=0$, then the metric \eqref{metric} is reduced to
\eq{
ds^2 = a^2\left(\frac{dt}{a} - dr\right)\left(\frac{dt}{a} + dr\right)
}{nullmetric}

We now define new coordinates
\eq{
u= \eta - r,~~  v= \eta + r
}{vu_def}
where $\eta = \int \frac{dt}{a}$ is a conformal time. Then we have $u=const$ for ongoing geodesics (in relation to the observer), and $v=const$ for ingoing ones. Since we are interested in ingoing geodesics, we choose $v=const$ or in terms of cosmic time $t$
\eq{
dr=-\frac{dt}{a}
}{dr}

The co-vector field
\eq{
k^{in}_\alpha = - \p_\alpha v
}{vfield}
should be a null one and has to satisfy the geodesics equation. From the metric \eqref{metric} it is easy to find
\eq{
k^{in}_\alpha = (-\frac{1}{a},-1,0,0),~~ k^\alpha_{in} = \left(-\frac{1}{a}, \frac{1}{a^2},0,0\right)
}{kalpha}

It is easy to check that $k^\alpha k_\alpha =0$.

Now we will find the affine parameter for $k^\alpha_{in}$ from the geodesic equation
\eq{
k^\alpha_{;\beta}k^\beta = \frac{\p k^\alpha}{\p \lambda} + \cf{\alpha}{\gamma\beta}k^\gamma k^\beta = 0
}{nullgeod}

By substituting \eqref{kalpha} to \eqref{nullgeod} we obtain
\eq{
%\frac{1}{a^2}\frac{\p a}{\p \lambda} + a\dot{a}\frac{1}{a^2}\frac{1}{a^2} =
\frac{\p a}{\p \lambda} + \frac{\dot{a}}{a} = 0
}{afinnpar_prep}
From the equation \eqref{afinnpar_prep} we can find the relation for the affine parameter
\eq{
d \lambda = - adt
}{lambda}

Using the obtained relation \eqref{lambda} we can calculate the expansion $\Theta $ \cite{Poisson_2004}:

\eq{
	\Theta \equiv k^\alpha_{;\alpha}
= - \frac{2}{a^2}\left(\dot{a} - \frac{f'(r)}{f(r)}\right)
}{theta2}

We can also use the geometrical interpretation of the expansion $\Theta$
\eq{
\Theta = \frac{1}{S}\frac{\p S}{\p \lambda}
}{thetaint}
It's easy to see that a congruence's cross-sectional area $S$ in the chosen metric \eqref{metric} is
\eq{
	S = \frac{\pi z^2}{4}
}{area}
where $z$ is the diameter of the cross-section.

Our task now is to represent  $\Theta$ through $z$ - the distance between two neighbour rays (one from the beginning of the object and another from the end of the object). Further, we will interpret $z$ as the diameter of the cross-section if we assume rotational symmetry in the FLRW metric. For the sake of simplicity, let the first light ray propagate along the axis $\phi=0$. As we noted at the beginning of this section, we firstly want to find a differential equation for $z$ in the Friedmann Universe. From the definition of the ADD in the curved space-time (for FLRW metric \eqref{metric} $d_a= a f(r)$) we have the distance between the two rays
\eq{
z = a\phi f(r)
}{z}

Substituting \eqref{z} and \eqref{area} into \eqref{thetaint} one can find
$
\Theta = -\frac{2\dot{z}}{az}
$
and using \eqref{theta2} we obtain

\eq{
\dot{z} = \frac{z}{a}\left(\dot{a} - \frac{f'(r)}{f(r)}\right)
}{dot-z}

For the derivative along the path, using \eqref{dr}, we find
\eq{
	\ddot{z}
= \frac{z}{a}\left(\ddot{a}-\frac{\dot{a}}{a}\frac{f'(r)}{f(r)} + \frac{f''(r)}{f(r)a}\right)
}{ddotz}

We can express $\frac{f'(r)}{f(r)}$ from \eqref{dot-z} and insert it into \eqref{ddotz}.
Then taking into account that $\frac{f''(r)}{f(r)}=k$, we obtain the equation
for an arbitrary curvature
\eq{
\ddot{z} - \frac{\dot{a}}{a}\dot{z} - \left(\frac{\ddot{a}}{a}-\frac{\dot{a}^2}{a^2} - \frac{k}{a^2}\right)z = 0
}{zeldovich_gen}

The initial conditions are presented in the following way
\eq{
z(t_0) = 0,\ \ \ \dot{z}|_{t_0} = \phi
}{ics}

The initial conditions are derived from the fact that the distance at the point of observation between two rays from the object is equal to zero, and the change in the velocity of this distance equals the angle of observation by definition, and the fact that $c=1$.

Our next task is to derive the expression for Ricci focusing in the Friedmann Universe.
To this end let us approach the work in \cite{Zeldovich_1965}, and restore the result for the sake of completeness without essential changes.

Let us define the angle $\psi$ between the rays by the formula
\eq{
	\tan \psi \cong \psi = -\frac{d z}{dt}
= -\phi f(r)\dot{a} + \phi f'(r)
}{psigeom}
where \eqref{z} and \eqref{dr} were used. The first ray propagates along the axis of the coordinates. Let us write the angle $\psi$ as the ratio of the perpendicular component of the momentum of the second photon $q$ to its longitudinal component $h$
\eq{
\psi = -\frac{q}{h}
}{psimom}
Since $|q|\leq h $ the total momentum $P$ is proportional to the frequency of the quantum, and it is equal to $h$. From the redshift formula, the total momentum $P$ is given by
\eq{
P = h = \hbar \omega = \frac{K}{a(t)}
}{tmomentum}
where $K= \hbar \omega(t_0)a(t_0)$ is a constant. Form \eqref{tmomentum} we obtain the expression for the transverse component $q$ in the case of propagation in the homogeneous universe
$$
q = -\psi h = \frac{K\phi}{a}\left(\dot{a}f(r)- f'(r)\right)
$$
while for the derivative of this component along the path, using \eqref{dr}, we find for arbitrary curvature $k=\{-1,0,1\}$
\eq{
\frac{d q}{dt} =  hz\left[\frac{\ddot{a}}{a} - \frac{\dot{a}^2}{a^2} - \frac{k}{a^2}\right]
}{difq_gen}

Thus we can rewrite \eqref{zeldovich_gen} using \eqref{difq_gen}
\eq{
%\boxed{
\ddot{z} - \frac{\dot{a}}{a}\dot{z} - \frac{\dot{q}}{h} = 0
%}
}{zeldovich_gen2}

Let us reaffirm ourselves that we derived the general differential equation for the Friedmann Universe where the part responsible for Ricci focusing is represented by the separate term.
We want to focus attention on the difference between the resulting differential equation and that obtained by V. Dashevsky and V. Slysh \cite{Dashevskii_1965}. The result obtained by them is valid for the special case -- a Friedmann Universe with $\Omega_0=\Omega_M=1$. Our equation \eqref{zeldovich_gen2} is valid for any type of Friedmann Universe: open, spatially-flat or closed model, and for any expression and form of the energy momentum tensor. Thus our equation is valid even for modified theories of gravitation if a Friedmann Universe is involved.

\section{Method of ADD calculation}
\label{sec:2}
%3. Method of ADD calculation

Let us investigate "homogeneous" and "homogeneous in the mean universe". We will study a Friedmann Universe filled by a perfect fluid (matter, radiation) with vacuum energy. We will consider the method of the ADD calculation for such a universe.

Let us start with the homogeneous universe.
The dynamics of the universe can be described by the Einstein-Friedmann equations
\eq{
\frac{\dot{a}^2}{a^2}+\frac{k}{a^2} = \frac{8\pi G}{3}\rho
}{friedman2}
\eq{
\frac{\ddot{a}}{a} = -\frac{4\pi G}{3} (3p+\rho)
}{friedman1}
Using these equations in \eqref{difq_gen} we can obtain the important relation
\eq{
\frac{\dot{q}}{h} = -4\pi G z(p+\rho)
}{qhhom}
Thus the equation \eqref{zeldovich_gen2} transforms to
\eq{
\ddot{z} - \frac{\dot{a}}{a}\dot{z} + 4\pi G z(p+\rho)= 0
}{homzeld}

We can find the solutions for \eqref{homzeld} independent of the form of the scale factor $a(t)$ and compatible with the initial conditions \eqref{ics}
\eq{
z = \left\{\begin{array}{lr}
	\phi a \sin\int_t^{t_0}\frac{dt}{a} & : k=1 \\
	\phi a \int_t^{t_0}\frac{dt}{a} & : k=0 \\
	\phi a \sh\int_t^{t_0}\frac{dt}{a} & : k=-1
\end{array}\right.
}{homansatz}
One can easily check these solutions by direct substitution of the solutions \eqref{homansatz} into \eqref{homzeld}.
Thus \eqref{homansatz} is the solution for the homogeneous universe, and the angular diameter distance can be calculated with the formulae:
\eq{
d_a = \left\{\begin{array}{lr}
	a \sin\int_t^{t_0}\frac{dt}{a} & : k=1 \\
	a \int_t^{t_0}\frac{dt}{a} & : k=0 \\
	a \sh\int_t^{t_0}\frac{dt}{a} & : k=-1
\end{array}\right.
}{dahom}

To obtain the dependence $d_a$ on redsift $f$\footnote{It shoud not be confused with f(r) used in the section 2.}, an expression of the form $d_a = \psi(f)$ for the most general cosmological model (for a mixture of vacuum energy and relativistic and non-relativistic matter) should involve the equation of state \cite{Weinberg_2008}
\eq{
-\frac{d\rho}{\rho + p} = 3d\ln a
}{eqofstate}
This equation can be solved for matter ($p=0$), radiation ($p=\rho/3$), and vacuum energy ($p_\Lambda=-\rho_\Lambda=const$). The expression for a mixture of them is
\eq{
\rho = \frac{3H_0^2}{8\pi G}\left[\Omega_\Lambda + \Omega_M\left(\frac{a_0}{a}\right)^3 + \Omega_R\left(\frac{a_0}{a}\right)^4\right]
}{rho}
where the present energy densities in the vacuum, non-relativistic matter, and relativistic matter are, respectively
\eq{
\rho_\Lambda = \frac{3H_0^2\Omega_\Lambda}{8\pi G},\ \ \rho_M = \frac{3H_0^2\Omega_M}{8\pi G},\ \ \rho_R = \frac{3H_0^2\Omega_R}{8\pi G}.
}{rholambda}
Using \eqref{rholambda}, Friedmann equation \eqref{friedman2} can be represented in the form
\eq{
dt = \frac{dx}{H_0x\sqrt{\Omega_S}}
}{dt}
where $x=\frac{a}{a_0} = \frac{1}{1+f}$ ($t=t_0 \Rightarrow x = 1$), $\Omega_k=-\frac{k}{a_0^2H_0^2}$,$f$ -redshift and
$$
\Omega_S = \Omega_\Lambda + \Omega_kx^{-2} + \Omega_M x^{-3} + \Omega_Rx^{-4}
$$

Thus following \eqref{dahom} we can present the resulting formulae for ADD in the form:
\eq{
d_a = \left\{\begin{array}{lr}
	\frac{1}{1+f}\frac{1}{H_0\sqrt{\Omega_k}} \sin\int_\frac{1}{1+f}^1\sqrt{\Omega_k}\frac{dx}{x^2\sqrt{\Omega_S}} & : k=1 \\
	\frac{1}{1+f} \int_\frac{1}{1+f}^1\frac{dx}{H_0x^2\sqrt{\Omega_S}} & : k=0 \\
	\frac{1}{1+f} \frac{1}{H_0\sqrt{\Omega_k}}\sh\int_\frac{1}{1+f}^1\sqrt{\Omega_k}\frac{dx}{x^2\sqrt{\Omega_S}} & : k=-1 \\
\end{array}\right.
}{DaHomZ}
Let us mention that the result \eqref{DaHomZ} is in agreement with widely used formulae represented in \cite{Hogg_2000}.

We should mention that \eqref{dahom} is valid for any modified gravitational theory where the "Einstein-Friedmann" equations can be written in the form

\eq{
\frac{\dot{a}^2}{a^2}+\frac{k}{a^2} = \psi(t)
}{mfriedman2}
\eq{
\frac{\ddot{a}}{a} = \xi(t)
}{mfriedman1}
Here $\xi(t)$ and $\psi(t)$ are arbitrary functions.

Let us turn our attention to the investigation of "homogeneous in the mean universe".

The description of such a universe contains two propositions: (1) we have a homogeneous distribution of matter along the whole universe; (2) the interaction between matter and light rays is negligible under standard observations.

The proposition (2) means that for this type of universe, we have $\dot{q}=0$ and equation \eqref{zeldovich_gen2} takes the form
\eq{
\ddot{z} - \frac{\dot{a}}{a}\dot{z} = 0
}{zeldovich_uhom}

The solution of \eqref{zeldovich_uhom} can be presented in a form compatible with the initial conditions \eqref{ics}
\eq{
z = \frac{\phi}{a_0}\int a dt
}{ansatz_uh}
Thus, the expression for the angular diameter distance for any curvature in the "homogeneous in the mean universe" is
\eq{
d_a = \frac{1}{a_0}\int_t^{t_0} a dt
}{dA_uhomc}
Note that the result is valid for any modified gravity theory when the Friedmann Universe is under consideration.

If we consider the $\Lambda$CDM model, the ADD formula \eqref{dA_uhomc}, with the help of \eqref{dt}, transforms to
\eq{
d_a = \int_\frac{1}{1+f}^1\frac{dx}{H_0\sqrt{\Omega_\Lambda + \Omega_kx^{-2} + \Omega_M x^{-3} + \Omega_Rx^{-4}}}
}{dA_uhom}

The obtained expression for ADD coincides with the solution found by V. Dashevskii and Ya. Zeldovich  \cite{Zeldovich_1964}, \cite{Zeldovich_1965}, when $\Omega_\Lambda = \Omega_R =0$.

\section{Numerical solutions}
\label{sec:3}
We now present numerical solutions to equation \eqref{zeldovich_gen2} for the partly filled cone for the present day Univese: $k=0$, $\Omega_R = 0$, $\Omega_M + \Omega_\Lambda = 1$.

As a first step, we may rewrite the Einstein-Friedmann equations \eqref{friedman1} and \eqref{friedman2} for the spatially-flat universe
\begin{eqnarray}\label{friedman_flat}
\frac{\ddot{a}}{a} = -\frac{4\pi G}{3}(\rho + 3p) \\
\frac{\dot{a}^2}{a^2} = \frac{8\pi G \rho}{3}
\end{eqnarray}
Let us write the solution for the equation of state \eqref{eqofstate} of a mixture of matter and vacuum energy
\eq{
\rho = \frac{3H_0^2}{8\pi G}\left[\Omega_\Lambda + \Omega_M\left(\frac{a_0}{a}\right)^3\right]
}{rhovacmat}
Now we make the substitution of \eqref{rhovacmat} into \eqref{friedman_flat}
\eq{
\frac{\dot{a}^2}{a^2}=H_0^2\left[\Omega_\Lambda + \Omega_M\left(\frac{a_0}{a}\right)^3\right]
}{subst1}
The solution (it is also presented in \cite{Mukhanov_2005} as the exercise) is
\eq{
a = a_0\left[\sqrt{\frac{\Omega_M}{\Omega_\Lambda}}\sinh\frac{3}{2}H_0\sqrt{\Omega_\Lambda}t\right]^\frac{2}{3}
}{fltsol}

\begin{figure*}[!htb]
\centering
\includegraphics[width=0.75\textwidth]{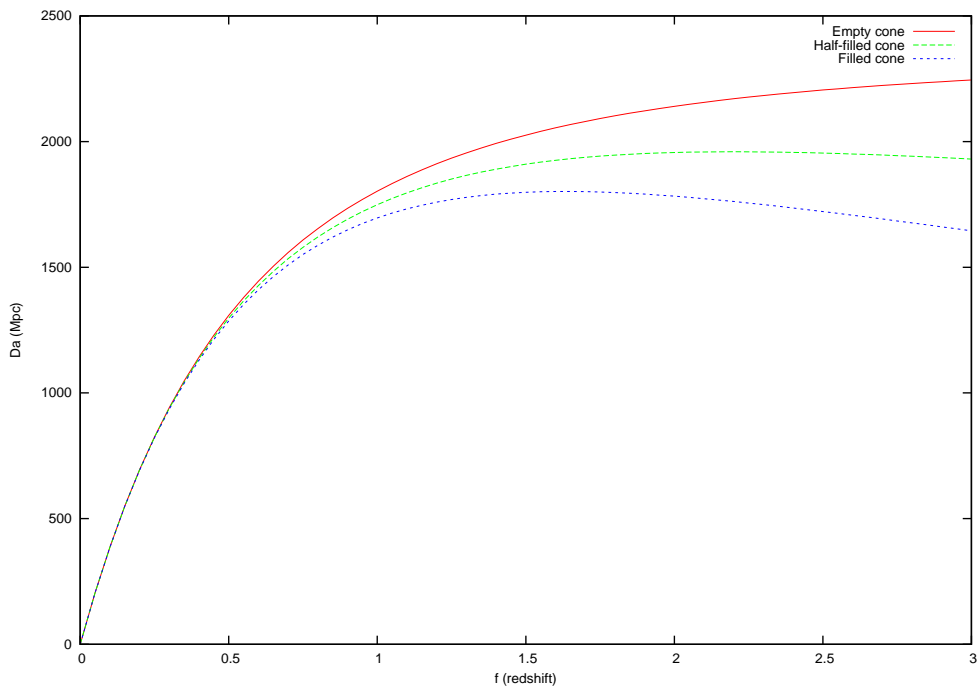}
\caption{The angular diameter distance with respect to redshift for Empty, Half-filled and Filled cones}
\label{figure2}
\end{figure*}

 We introduce $\alpha$ as the coeficient of the light cone "fillness", i.e. $\alpha = 1$ for the filled cone (with the critical density) and $\alpha = 0$ for the empty cone (with null density inside). Thus we can rewrite our equation \eqref{zeldovich_gen} in the form
\eq{
\ddot{z} - \frac{\dot{a}}{a}\dot{z} - \alpha\frac{\dot{q}}{h} = 0
}{alphazeld}
Using \eqref{fltsol} we can transform \eqref{alphazeld} to
\eq{
\ddot{z} - k\coth(\frac{3}{2}kt)\dot{z} + \frac{3}{2}\alpha k^2\left(\sinh\frac{3}{2}kt\right)^{-2}z = 0
}{zeld_z}
with initial conditions $z(t_0) = 0$ and $\left.\dot{z}\right|_{t_0}=\phi $. We know that $d_a =\frac{z}{\phi}$, thus we can rewrite \eqref{zeld_z}
\eq{
\ddot{d_a} - k\coth(\frac{3}{2}kt)\dot{d_a} + \frac{3}{2}\alpha k^2\left(\sinh\frac{3}{2}kt\right)^{-2}d_a = 0
}{zeld_da}
with initial conditions $d_a(t_0) = 0$ and $\left.\dot{d_a}\right|_{t_0}=1$. In terms of the redshift $f$ we may calculate $t_e$ - the time of emission and $t_0$ - the time of observation with the relation
\eq{
t  = \frac{1}{h_0}\int_0^\frac{1}{1+f}\frac{dx}{x\sqrt{\Omega_\Lambda + \Omega_M x^{-3}}}
}{teq}
for $t_e$. For $t_0$ we should select $f=0$. The results are shown in Figure \ref{figure2}. It is clear from the plots in Figure \ref{figure2} that the difference in the ADD measurement may lead to 600-700 Mps at $f \simeq 3$.

The GNU Octave code\footnote{https://www.gnu.org/software/octave/} for solving \eqref{zeld_da} may be found on the web-page http://lgca.ulspu.ru/nikolaev.

\section{Conclusions}
We extended Ya. Zeldovich's ideas for ADD measurements in two directions. Firstly, the generalization of the ADD formula from a closed to spatially-flat and open Friedmann Universes. Secondly, we proposed not only empty and filled cone of light rays (CLR), but also the partly filled CLR.

These main results are represented by the differential equation \eqref{zeldovich_gen2} which allows us to separate the effects of expansion of the homogeneous universe and Ricci focusing for congruence of radial null geodesics. It was shown, that this equation consists of a classical solution for the  homogeneous Friedmann Universe and, as a special case, it reduces to the equations obtained by Zeldovich \etal  The solution of \eqref{zeldovich_gen2} was presented in quadratures in a form suitable for further numerical analysis.

The numerical solution for the partly filled CLR was obtained. From this solution (Figure 2) it became evident that the standard ADD measurement (in the Universe filled by (dark and baryonic) matter and dark energy) may be applied for an object with redshift $f$ of not more than 0.5. For objects with $f>0.5$, the influence of CLR filling became crucial. For example, for a redshift of $f \simeq 3$, the ADD may have a difference of about 600-700 Mps. for empty and totally filled CLR.

These results can help astronomers to improve their calculations where ADD is involved. We plan to present an extension of this method to gravitational lensing, supernova data analysis in the next publication.

\begin{acknowledgements}
AVN would like to thank Robert Schmidt for very useful discussions, and Joachim Wambsganss for support. Also, AVN would like to thank ARI\footnote{Astronomisches Rechen-Institut} of ZAH\footnote{Zentrum fuer Astronomie der Universität Heidelberg} for useful discussions at the seminars on astronomy. Aleksei Nikolaev is grateful for the grant from the Ministry of education and science of RF, named in the honour of the president of the Russian Federation, for supporting his scholarship at ZAH where this work was initiated.

Authors are thankful to the referee, who attracted their attention to the investigation of inhomogeneities and its connection to a luminosity distance in a series of works.

SVC and AVN were supported by the State order of Ministry of education and science of RF number 2014/391 on the project 1670.
\end{acknowledgements}

\section{Comment}
The present remarks to this article result from
useful discussion with P. Helbig, one of the authors of the paper \cite{Kayser_1996}. We would like to clarify this in what follows.

While it is true that Eq. (24) in Kayser et al. (1997)\cite{Kayser_1996} is essentially
the same as Eq. (2) in Dashevskii \& Slysh (1966)\cite{Dashevskii_1965}, which was derived
assuming that the cosmological constant is zero, and that the more
general equation is our Eq. (32), which takes all equations of state
into account, it is nevertheless the case, contrary to our original
claim, that the equations in Kayser et al. (1997)\cite{Kayser_1996} are correct.  The
reason for this is that in the special case of the cosmological
constant $w = -1$ and hence $p = -\rho$, so the additional terms for
pressure and energy density cancel each other.  (This is not the case
for other equations of state, but Kayser et al. (1997)\cite{Kayser_1996} do not claim
to handle these; they consider only non-relativistic matter, the
cosmological constant, and curvature.  For all of these, and for the
homogeneity parameter, all physically meaningful values are allowed.)

The another derivation of general equation for ADD measurement and the detail proof of the fact that dark energy in standard form do not affect focusing are presented in paper \cite{Nikolaev_2016_gc}.

\bibliographystyle{spphys}       % APS-like style for physics
\bibliography{Nikolaev_literature}   % name your BibTeX data base

%\begin{thebibliography}{}
% etc
%\end{thebibliography}

\end{document}